\documentclass[12pt]{article}
\usepackage{graphicx}

\newsavebox{\sboxpubnumber}
\newsavebox{\sboxpubdate}
\newcommand{\sss}{\scriptscriptstyle}
\newcommand{\beq}{\begin{equation}}
\newcommand{\eeq}{\end{equation}}
\newcommand{\beqa}{\begin{eqnarray}}
\newcommand{\eeqa}{\end{eqnarray}}
\newcommand{\pubdate}[1]{\begin{lrbox}{\sboxpubdate}{#1}\end{lrbox}}
\newcommand{\pubnumber}[1]{\begin{lrbox}{\sboxpubnumber}{\begin{tabular}{l} #1 \\
				 \usebox{\sboxpubdate}
				 \end{tabular}}
                           \end{lrbox}
                           \pubblock}
\newcommand{\Title}[1]{\begin{center} {\Large #1 } \end{center}}
\newcommand{\Author}[1]{\begin{center}{ \sc #1} \end{center}}
\newcommand{\Address}[1]{\begin{center}{ \it #1} \end{center}}

\newcommand{\pubblock}{\rightline{
			\usebox{\sboxpubnumber}}}
\newenvironment{Abstract}{\begin{quotation}  }{\end{quotation}}
\newenvironment{Presented}{\begin{quotation} \begin{center}
             PRESENTED AT\end{center}\bigskip
      \begin{center}\begin{large}}{\end{large}\end{center}
      \end{quotation}}

\begin{document}

%%%%%%%%%%%%%%%%%%%%%%%%%%%%%%%%%%%%%%%%%%%%%%%%%%%%%%%%%%%%%%%%%%%%%%%%
%%
%% START EDITING HERE!
%%
%%%%%%%%%%%%%%%%%%%%%%%%%%%%%%%%%%%%%%%%%%%%%%%%%%%%%%%%%%%%%%%%%%%%%%%%
\begin{titlepage}
\pubdate{\today}                    %fill in the date
\pubnumber{McGill-02-02} %preprint number(s)

\vfill
\Title{Electroweak Phase Transition and Baryogenesis}
\vfill
\Author{James M.\ Cline}
\Address{Physics Department, McGill University\\
         3600 University St., Montr\'eal, Qu\'ebec, Canada H3A 2T8}
\vfill
\begin{Abstract}
I review the status of the strength of the electroweak phase transition,
and electroweak baryogenesis in the Minimal Supersymmetric Standard Model
(work done with K.\ Kainulainen and M.\ Joyce).  The emphasis is on new
brane-inspired ideas about electroweak baryogenesis, and improvements in the
semiclassical treatment of CP violation at a first order electroweak phase
transition.
\end{Abstract}
\vfill
\begin{Presented}
    COSMO-01 \\
    Rovaniemi, Finland, \\
    August 29 -- September 4, 2001
\end{Presented}
\vfill
\end{titlepage}
\def\thefootnote{\fnsymbol{footnote}}
\setcounter{footnote}{0}

%%%%%%%%%%%%%%%%%%%%%%%%%%%%%%%%%%%%%%%%%%%%%%%%%%%%%%%%%%%%%%%%%%%%%%%%
% The document starts here
%%%%%%%%%%%%%%%%%%%%%%%%%%%%%%%%%%%%%%%%%%%%%%%%%%%%%%%%%%%%%%%%%%%%%%%%
\section{Introduction}
I have given a similar talk in ref.\ \cite{whepp} in 2000, so I will not repeat
material that was presented there.  Rather I will emphasize what is new since
that time.  The main new results fall under the categories of (1) effects of
brane cosmology, (2) nonthermal production of sphalerons by preheating, and
(3) refinements of the computation of electroweak baryogenesis in the MSSM.

\section{Brane-world implications for Baryogenesis}

In the context of large (ADD \cite{ADD}) extra dimensions, it has already
been noted that baryogenesis is difficult because of the extremely low reheat
temperature that is needed to keep light Kaluza-Klein gravitons out of
thermal equilibrium, since they would distort the cosmic gamma ray background
\cite{BD}.  A warped extra dimension {\it \`a la} Randall and Sundrum (RS)
\cite{RS} however provides an interesting possibility: the Friedmann equation
is modified to the form \cite{BDL}-\cite{CGS}
\beq
\label{feq}
	H^2 = {8\pi G\over 3}\rho\left(1+{\rho\over\Lambda}\right)
\eeq
where $\Lambda$ is the tension of the brane on which we are presumed to be living.
If $\Lambda$ is sufficiently small, then the expansion rate could be significantly
increased at the time of the electroweak phase transition \cite{Servant}, making
it possible for sphalerons to go out of equilibrium as is required for electroweak
baryogenesis, without having to add new physics for the purpose of strengthening
the electroweak phase transition.  Unfortunately the modified Friedmann equation
(\ref{feq}) is specific to the RS-II model in which there is a just a single
brane; in the two-brane version that was invented to solve the hierarchy problem,
the $O(\rho^2)$ correction has the wrong sign for helping with baryogenesis
\cite{CV}.  It would be interesting to find other brane-world models which had
the desired behavior.

On a more general note, it could be expected that cosmology at the TeV scale
(or perhaps 100 GeV if we push the parameters) might be rather radically 
altered in the RS-I model, since the TeV brane on which we live in that model
should cease to exist as we know it at temperatures exceeding this scale
\cite{APR}-\cite{CNR}.  At sufficiently high temperatures the TeV brane is
hidden behind a horizon in the extra dimension.  Its emergence is associated with
a phase transition in the conformal field theory corresponding to the bulk graviton
degrees of freedom.  If this coincides with the electroweak phase transition,
the situation could be richer than is normally assumed.

\section{Preheating Effects}

One of the holy grails in electroweak baryogenesis is satisfying the  sphaleron
bound: the rate of sphaleron interactions per unit volume  must be less than $H^4$
once the baryon asymmetry has been created in  order to avoid its relaxation back
to negligible levels.  I mentioned eq.\ (\ref{feq}) as one new idea for achieving
this.  Another which has gotten some attention recently is related to preheating. 
If inflation occurs with a low reheat temperature $T\ll 100$ GeV (as one would
like for the large extra dimension scenarios), then sphalerons could not be
created by normal thermal processes.  However they might be produced by
nonequilibrium conversion of the coherent field energy of the inflaton  near
the end of inflation, as has been discussed by A.\ Rajantie in these proceedings
\cite{Rajantie} and in the references \cite{nonequil}.  

In order to make this idea work, one can couple the inflaton to the Higgs 
field in the manner of hybrid inflation, through an interaction of the 
form
\beq
	V= \lambda(H^2-v^2)^2 + \frac12 m_\sigma^2 \sigma^2 + g^2\sigma^2 H^2
\eeq
but the COBE observations require that $m_\sigma\sim 10^{-10}$ eV, which is
unstable against radiative corrections \cite{Lyth}.  The problem can be
ameliorated in inverted hybrid inflation models\cite{CLRT}, in which the inflaton
$\sigma$ rolls away from $\sigma=0$ instead of towards it, due to the addition of
some nonrenormalizable operators to $V$.  The number of counterterms which need
to be fine tuned to keep $V$ sufficiently flat in this model is much smaller than
in the ordinary hybrid model.  Nevertheless, this discussion underscores the
difficulties in constructing a convincing or natural realization of low temperature
baryogenesis.

\section{Electroweak Baryogenesis in the MSSM}

It has become standard to compute the baryon asymmetry due to sphalerons in three
steps \cite{HN}-\cite{HJS}: first compute the source term that appears in
diffusion equations for the various species which couple to left-handed quarks
(since these are the particles which ultimately bias sphalerons); then solve the 
diffusion equations for the chemical potentials of the left-handed quarks. This is
easily fed into the sphaleron rate equation to compute the baryon asymmetry.
Because the network of diffusion equations is complicated, it was also standard
to employ some simplifying approximations, one of which was to assume that the sum
of the chemical potentials for the two Higgs fields $H_1$ and $H_2$ is driven
to zero by interactions involving the top quark Yukawa coupling.  In this
approximation one considers only the source for the difference between the 
two, and this turns out to be suppressed by the fact that the ratio $H_1/H_2$
remains quite constant within the bubble walls that form during the electroweak
phase transition \cite{MQS,CM}.  However, it was pointed out by us \cite{CK}
that the assumption of top Yukawa equilibrium is not realistic, and that the
source term for $H_1+H_2$ is much larger than that for $H_1-H_2$, since it is
not suppressed by the constancy of $H_1/H_2$ inside the wall, nor is it very
strongly suppressed by the top Yukawa interactions.

Despite this enhancement, we still find that the baryon asymmetry is small, and it
is rather difficult to tune the parameters of the MSSM to get an acceptably large
baryon asymmetry.  Our results are in contrast to those of \cite{CQRVW, CMQSW,
RSanz}, who find larger values.  The difference comes from our respective
derivations of the source term in the diffusion equations, which we do starting
from the semiclassical CP-violating force acting on Higgsinos in the bubble
wall \cite{CJK,CJK2}.  From this force, one can derive the diffusion equations
from the Boltzmann equation in a controlled and rigorous fashion.

It should be emphasized that if we were simply making rough order-of-magnitude
estimates of the baryon asymmetry based on the semiclassical formalism, our results
would be in better agreement with those of \cite{CQRVW, CMQSW, RSanz}.  The
suppression comes from the detailed properties of our source term for the
Higgsinos, which looks like
\beq
	S_{\sss H} = {v_w D_{\sss \widetilde H}\over 2\langle v^2\rangle T}
	\left\langle{|p_z|\over E^3}\gamma\right\rangle \left(m^2\theta'\right)''
\eeq
where $v_w$ is the bubble wall velocity, $D_{\sss \widetilde H}$ is the Higgsino
diffusion coefficient, $\langle \cdot \rangle$ denotes thermal averaging,
and $m e^{i\theta(x)}$ is the locally varying Higgsino mass inside the wall.
The salient feature is that this source is close to being a total derivative,
and it must be integrated in the solution of the diffusion equations.  Its
integral is much smaller than its typical values, which we would have used 
had we been doing an order-of-magnitude estimate.  This can be seen from 
Fig.\ 1(a), which shows $S_{\sss H}$ as a function of distance in the wall,
and Fig.\ 1(b), the left-handed quark chemical potential.  The latter is several 
orders of magnitude smaller than the former due to the large cancellations
which take place in integrating the diffusion equations.
\vspace{-0.01in}
%%%%%%%%%%%%%%%%%%%%%%%%%%%%%%%%%%%%%%%%%%%%%%%%%%%%%%%%%%%%%%%%%%%%%%%%
%%
%%   use this format to include an .eps figure into your paper
%%
\begin{figure}[h]
    \centering
    \includegraphics[height=3.15in]{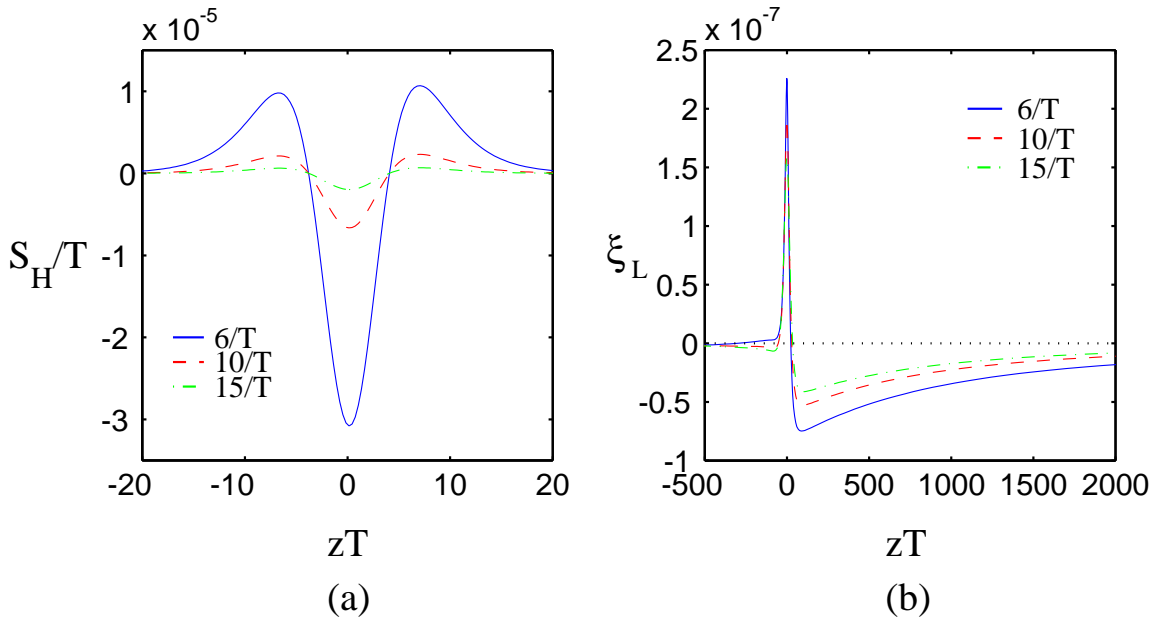}
    \caption{(a) Higgsino source term versus distance $\times$ temperature in wall. (b) Left-handed quark
chemical potential over $T$, versus distance $\times T$.}
    \label{fig:cosmo}
\end{figure}
%%%%%%%%%%%%%%%%%%%%%%%%%%%%%%%%%%%%%%%%%%%%%%%%%%%%%%%%%%%%%%%%%%%%%%%%

As a result, we are forced to take the CP violating phase of the $\mu$
parameter to be close to maximal, and to assume that the wall velocity
$v_w$ and $\tan\beta$ are close to their optimal values, as shown in fig.\
2.  The required value of $v_w$ is not unlikely \cite{wall}, but
such large phases require the squarks to be quite heavy in order to
suppress the loop contributions to the EDM of Mercury \cite{AKL}.
Moreover, we need to take the chargino and Higgsino mass parameters
$|\mu|$ and $m_2$ to be nearly degenerate, as shown in fig.\ 3.

%%%%%%%%%%%%%%%%%%%%%%%%%%%%%%%%%%%%%%%%%%%%%%%%%%%%%%%%%%%%%%%%%%%%%%%%
%%
%%   use this format to include an .eps figure into your paper
%%
\begin{figure}[h]
    \centering
    \includegraphics[height=3.15in]{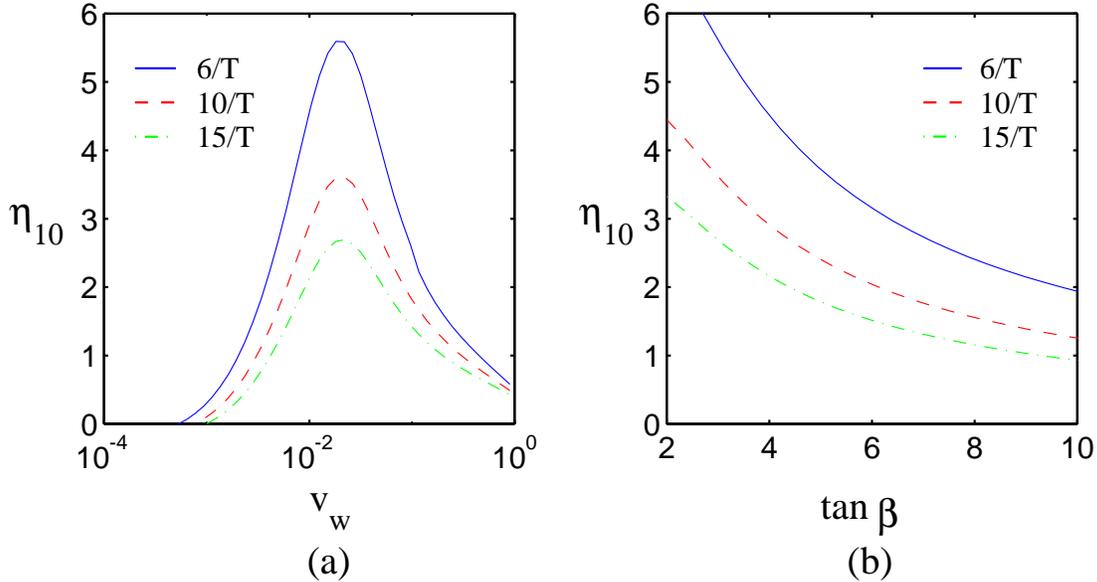}
    \caption{Baryon-to-photon ratio $\times 10^{10}$ as a function of 
(a) wall velocity and (b) $\tan\beta = \langle H_2\rangle/\langle H_1\rangle$.
}
    \label{fig:cosmo2}
\end{figure}
%%%%%%%%%%%%%%%%%%%%%%%%%%%%%%%%%%%%%%%%%%%%%%%%%%%%%%%%%%%%%%%%%%%%%%%%
%%%%%%%%%%%%%%%%%%%%%%%%%%%%%%%%%%%%%%%%%%%%%%%%%%%%%%%%%%%%%%%%%%%%%%%%
%%
%%   use this format to include an .eps figure into your paper
%%
\begin{figure}[h]
    \centering
    \includegraphics[height=3.15in]{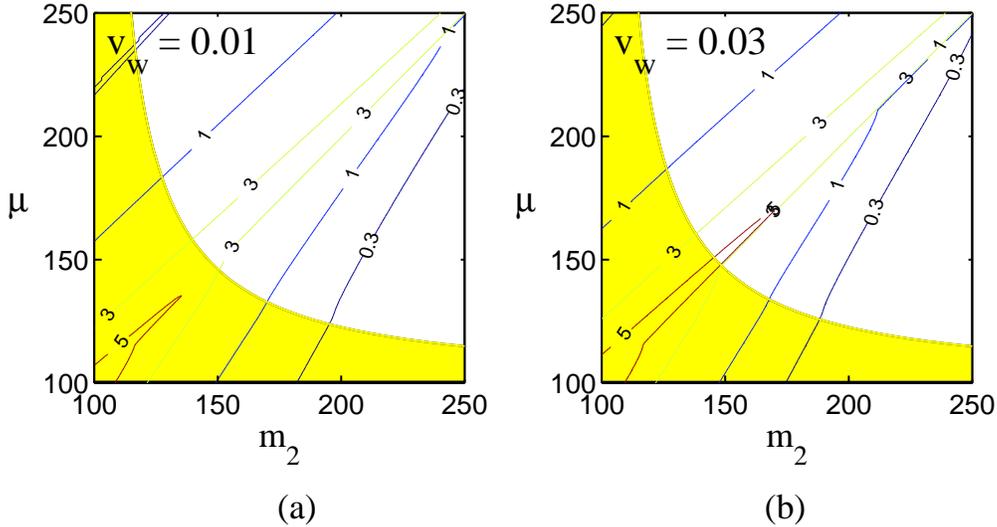}
    \caption{Baryon-to-photon ratio contours in the plane of chargino mass
parameters for two different wall velocities.  Shaded regions are excluded
by LEP II.}
    \label{fig:cosmo3}
\end{figure}
%%%%%%%%%%%%%%%%%%%%%%%%%%%%%%%%%%%%%%%%%%%%%%%%%%%%%%%%%%%%%%%%%%%%%%%%

In addition to the tunings of parameters already mentioned, one needs for the
right-handed stop mass to be very light \cite{CQW, CK0, CM} and the left-handed
stop to be heavy.  Therefore electroweak baryogenesis in the MSSM is not yet ruled 
out, but it is close to being so.   Life is easier in the NMSSM though, where the
MSSM is supplemented by a singlet field.  Not only is it easy to make the
electroweak phase transition stronger \cite{nmssm}, but the CP violation can occur
transitionally and thus be relatively free from experimental constraints
\cite{HS}.  The semiclassical analysis of the source term in this model has been
carried out in \cite{kimmo}.


\begin{thebibliography}{99}

%%
%%  bibliographic items can be constructed using the LaTeX format in
%%  SPIRES:
%%  see    http://www.slac.stanford.edu/spires/hep/latex.html
%%  SPIRES will also supply the CITATION line information; please
%%  include it.
%%

\bibitem{whepp}
J.~M.~Cline,
``Status of electroweak phase transition and baryogenesis,''
Pramana {\bf 54}, 1 (2000)
[Pramana {\bf 55}, 33 (2000)]
[arXiv:hep-ph/0003029].
%%CITATION = HEP-PH 0003029;%%

\bibitem{ADD}\
{ N. Arkani-Hamed, S. Dimopoulos and G. Dvali,  Phys.\
Lett. {\bf B429} (1998) 263 [hep-ph/9803315]; Phys.\ Rev.\ {\bf D59}
(1999) 086004 [hep-ph/9807344]}

\bibitem{BD}
K.~Benakli and S.~Davidson,
``Baryogenesis in models with a low quantum gravity scale,''
Phys.\ Rev.\ D {\bf 60}, 025004 (1999)
[arXiv:hep-ph/9810280].
%%CITATION = HEP-PH 9810280;%%


\bibitem%[RS]
{RS}{L.~Randall and R.~Sundrum,
``A large mass hierarchy from a small extra dimension,''
Phys.\ Rev.\ Lett.\  {\bf 83}, 3370 (1999)
[hep-ph/9905221];
%%CITATION = HEP-PH 9905221;%%
L.~Randall and R.~Sundrum,
``An alternative to compactification,''
Phys.\ Rev.\ Lett.\  {\bf 83}, 4690 (1999)
[hep-th/9906064]
%%CITATION = HEP-TH 9906064;%%
}

\bibitem{BDL}
P.~Binetruy, C.~Deffayet and D.~Langlois,
``Non-conventional cosmology from a brane-universe,''
Nucl.\ Phys.\ B {\bf 565}, 269 (2000)
[hep-th/9905012].
%%CITATION = HEP-TH 9905012;%%

\bibitem{CGKT}
C.~Csaki, M.~Graesser, C.~Kolda and J.~Terning,
``Cosmology of one extra dimension with localized gravity,''
Phys.\ Lett.\ B {\bf 462}, 34 (1999)
[hep-ph/9906513].
%%CITATION = HEP-PH 9906513;%%

\bibitem{CGS}
J.~M.~Cline, C.~Grojean and G.~Servant,
``Cosmological expansion in the presence of extra dimensions,''
Phys.\ Rev.\ Lett.\  {\bf 83}, 4245 (1999)
[hep-ph/9906523].
%%CITATION = HEP-PH 9906523;%%

\bibitem{Servant}
G.~Servant,
``A way to reopen the window for electroweak baryogenesis,''
arXiv:hep-ph/0112209.
%%CITATION = HEP-PH 0112209;%%

\bibitem{CV}
J.~M.~Cline and J.~Vinet,
``Order $\rho^2$ corrections to Randall-Sundrum I cosmology,''
arXiv:hep-th/0201041.
%%CITATION = HEP-TH 0201041;%%

\bibitem{APR}
N.~Arkani-Hamed, M.~Porrati and L.~J.~Randall,
%``Holography and phenomenology,''
JHEP {\bf 0108}, 017 (2001)
[hep-th/0012148].
%%CITATION = HEP-TH 0012148;%%

\bibitem{HMR}
A.~Hebecker and J.~March-Russell,
%``Randall-Sundrum II cosmology, AdS/CFT, and the bulk black hole,''
Nucl.\ Phys.\ B {\bf 608}, 375 (2001)
[hep-ph/0103214].
%%CITATION = HEP-PH 0103214;%%

\bibitem{CNR}
P.~Creminelli, A.~Nicolis and R.~Rattazzi,
%``Holography and the electroweak phase transition,''
hep-th/0107141.
%%CITATION = HEP-TH 0107141;%%

\bibitem{Rajantie}
A.~Rajantie,
``Baryogenesis at the end of hybrid inflation,''
arXiv:hep-ph/0111200.
%%CITATION = HEP-PH 0111200;%%


\bibitem{nonequil}J.~Garcia-Bellido and A.~D.~Linde,
%``Preheating in hybrid inflation,''
Phys.\ Rev.\ D {\bf 57}, 6075 (1998)
[arXiv:hep-ph/9711360];
%%CITATION = HEP-PH 9711360;%%
J.~Garcia-Bellido, D.~Y.~Grigoriev, A.~Kusenko and M.~E.~Shaposhnikov,
%``Non-equilibrium electroweak baryogenesis from preheating after  inflation,''
Phys.\ Rev.\ D {\bf 60}, 123504 (1999)
[arXiv:hep-ph/9902449];
%%CITATION = HEP-PH 9902449;%%
L.~M.~Krauss and M.~Trodden,
%``Baryogenesis below the electroweak scale,''
Phys.\ Rev.\ Lett.\  {\bf 83}, 1502 (1999)
[arXiv:hep-ph/9902420];
%%CITATION = HEP-PH 9902420;%%
J.~M.~Cornwall and A.~Kusenko,
%``Baryon number non-conservation and phase transitions at preheating,''
Phys.\ Rev.\ D {\bf 61}, 103510 (2000)
[arXiv:hep-ph/0001058];
%%CITATION = HEP-PH 0001058;%%
J.~M.~Cornwall, D.~Grigoriev and A.~Kusenko,
%``Resonant amplification of electroweak baryogenesis at preheating,''
Phys.\ Rev.\ D {\bf 64}, 123518 (2001)
[arXiv:hep-ph/0106127];
%%CITATION = HEP-PH 0106127;%%
J.~Garcia-Bellido and D.~Y.~Grigoriev,
%``Inflaton-induced sphaleron transitions,''
JHEP {\bf 0001}, 017 (2000)
[arXiv:hep-ph/9912515];
G.~N.~Felder, J.~Garcia-Bellido, P.~B.~Greene, L.~Kofman, A.~D.~Linde and I.~Tkachev,
%``Dynamics of symmetry breaking and tachyonic preheating,''
Phys.\ Rev.\ Lett.\  {\bf 87}, 011601 (2001)
[arXiv:hep-ph/0012142];
%%CITATION = HEP-PH 0012142;%%%%CITATION = HEP-PH 9912515;%%
A.~Rajantie, P.~M.~Saffin and E.~J.~Copeland,
%``Electroweak preheating on a lattice,''
Phys.\ Rev.\ D {\bf 63}, 123512 (2001)
[arXiv:hep-ph/0012097].
%%CITATION = HEP-PH 0012097;%%

\bibitem{Lyth}
D.~H.~Lyth,
``Constraints on TeV-scale hybrid inflation and comments on non-hybrid  alternatives,''
Phys.\ Lett.\ B {\bf 466}, 85 (1999)
[arXiv:hep-ph/9908219].
%%CITATION = HEP-PH 9908219;%%

\bibitem{CLRT}
E.~J.~Copeland, D.~Lyth, A.~Rajantie and M.~Trodden,
``Hybrid inflation and baryogenesis at the TeV scale,''
Phys.\ Rev.\ D {\bf 64}, 043506 (2001)
[arXiv:hep-ph/0103231].
%%CITATION = HEP-PH 0103231;%%

\bibitem{HN}
P.~Huet and A.~E.~Nelson,
``Electroweak baryogenesis in supersymmetric models,''
Phys.\ Rev.\ D {\bf 53}, 4578 (1996)
[arXiv:hep-ph/9506477].
%%CITATION = HEP-PH 9506477;%%

\bibitem{CQRVW}
M.~Carena, M.~Quiros, A.~Riotto, I.~Vilja and C.~E.~Wagner,
``Electroweak baryogenesis and low energy supersymmetry,''
Nucl.\ Phys.\ B {\bf 503}, 387 (1997)
[arXiv:hep-ph/9702409].
%%CITATION = HEP-PH 9702409;%%

\bibitem{CJK}
J.~M.~Cline, M.~Joyce and K.~Kainulainen,
``Supersymmetric electroweak baryogenesis in the WKB approximation,''
Phys.\ Lett.\ B {\bf 417}, 79 (1998)
[Erratum-ibid.\ B {\bf 448}, 321 (1998)]
[arXiv:hep-ph/9708393].
%%CITATION = HEP-PH 9708393;%%

\bibitem{CJK2}
J.~M.~Cline, M.~Joyce and K.~Kainulainen,
``Supersymmetric electroweak baryogenesis,''
JHEP {\bf 0007}, 018 (2000)
[arXiv:hep-ph/0006119].
%%CITATION = HEP-PH 0006119;%%
erratum: arXiv:hep-ph/0110031.
%%CITATION = HEP-PH 0110031;%%

\bibitem{CMQSW}
M.~Carena, J.~M.~Moreno, M.~Quiros, M.~Seco and C.~E.~Wagner,
``Supersymmetric CP-violating currents and electroweak baryogenesis,''
Nucl.\ Phys.\ B {\bf 599}, 158 (2001)
[arXiv:hep-ph/0011055].
%%CITATION = HEP-PH 0011055;%%

\bibitem{HJS}
S.~J.~Huber, P.~John and M.~G.~Schmidt,
%``Bubble walls, CP violation and electroweak baryogenesis in the MSSM,''
Eur.\ Phys.\ J.\ C {\bf 20}, 695 (2001)
[arXiv:hep-ph/0101249].
%%CITATION = HEP-PH 0101249;%%



\bibitem{MQS}
J.~M.~Moreno, M.~Quiros and M.~Seco,
``Bubbles in the supersymmetric standard model,''
Nucl.\ Phys.\ B {\bf 526}, 489 (1998)
[arXiv:hep-ph/9801272].
%%CITATION = HEP-PH 9801272;%%

\bibitem{CM}
J.~M.~Cline and G.~D.~Moore,
``Supersymmetric electroweak phase transition: Baryogenesis versus experimental constraints,''
Phys.\ Rev.\ Lett.\  {\bf 81}, 3315 (1998)
[arXiv:hep-ph/9806354].
%%CITATION = HEP-PH 9806354;%%

\bibitem{CK}
J.~M.~Cline and K.~Kainulainen,
``A new source for electroweak baryogenesis in the MSSM,''
Phys.\ Rev.\ Lett.\  {\bf 85}, 5519 (2000)
[arXiv:hep-ph/0002272].
%%CITATION = HEP-PH 0002272;%%

\bibitem{RSanz}
N.~Rius and V.~Sanz,
``Supersymmetric electroweak baryogenesis,''
Nucl.\ Phys.\ B {\bf 570}, 155 (2000)
[arXiv:hep-ph/9907460].
%%CITATION = HEP-PH 9907460;%%

\bibitem{wall}
G.~D.~Moore,
``Electroweak bubble wall friction: Analytic results,''
JHEP {\bf 0003}, 006 (2000)
[arXiv:hep-ph/0001274].\\
%%CITATION = HEP-PH 0001274;%%
P.~John and M.~G.~Schmidt,
``Bubble wall velocity in the MSSM,''
arXiv:hep-ph/0012077.
%%CITATION = HEP-PH 0012077;%%

\bibitem{AKL}
T.~Falk, K.~A.~Olive, M.~Pospelov and R.~Roiban,
``MSSM predictions for the electric dipole moment of the Hg-199 atom,''
Nucl.\ Phys.\ B {\bf 560}, 3 (1999)
[arXiv:hep-ph/9904393].\\
%%CITATION = HEP-PH 9904393;%%
S.~Abel, S.~Khalil and O.~Lebedev,
``EDM constraints in supersymmetric theories,''
hep-ph/0103320.
%%CITATION = HEP-PH 0103320;%%

\bibitem{CQW}
M.~Carena, M.~Quiros and C.~E.~Wagner,
``Opening the Window for Electroweak Baryogenesis,''
Phys.\ Lett.\ B {\bf 380}, 81 (1996)
[arXiv:hep-ph/9603420].
%%CITATION = HEP-PH 9603420;%%

\bibitem{CK0}
J.~M.~Cline and K.~Kainulainen,
``Supersymmetric Electroweak Phase Transition: Beyond Perturbation Theory,''
Nucl.\ Phys.\ B {\bf 482}, 73 (1996)
[arXiv:hep-ph/9605235].
%%CITATION = HEP-PH 9605235;%%

\bibitem{nmssm}
A.~T.~Davies, C.~D.~Froggatt and R.~G.~Moorhouse,
``Electroweak Baryogenesis in the Next to Minimal Supersymmetric Model,''
Phys.\ Lett.\ B {\bf 372}, 88 (1996)
[arXiv:hep-ph/9603388].\\
%%CITATION = HEP-PH 9603388;%%
S.~J.~Huber,
``Singlets and the electroweak phase transition,''
arXiv:hep-ph/9902325.
%%CITATION = HEP-PH 9902325;%%

\bibitem{HS}
S.~J.~Huber and M.~G.~Schmidt,
``Electroweak baryogenesis: Concrete in a SUSY model with a gauge  singlet,''
Nucl.\ Phys.\ B {\bf 606}, 183 (2001)
[arXiv:hep-ph/0003122].
%%CITATION = HEP-PH 0003122;%%

\bibitem{kimmo}
K.~Kainulainen, T.~Prokopec, M.~G.~Schmidt and S.~Weinstock,
``First principle derivation of semiclassical force for electroweak  baryogenesis,''
JHEP {\bf 0106}, 031 (2001)
[arXiv:hep-ph/0105295].
%%CITATION = HEP-PH 0105295;%%



\end{thebibliography}
\end{document}